\documentstyle[preprint]{jpsj}

\newcommand{\vct}[1]{\mbox{\boldmath $#1$}}
\newcommand{\lsim}
 {\ \raise.35ex\hbox{$<$}\kern-0.75em\lower.5ex\hbox{$\sim$}\ }
\def\runtitle{
Theory of Local Density of States of 
$d_{x^{2}-y^{2}}$-Wave Superconducting State Near the 
Surfaces of the $t$-$J$ Model }
\def\runauthor{Yasunari {\sc Tanuma}, Yukio {\sc Tanaka}, 
Masao {\sc Ogata}, and Satoshi {\sc Kashiwaya}}

\title
{Theory of Local Density of States of 
$d_{x^{2}-y^{2}}$-Wave Superconducting State Near the 
Surfaces of the $t$-$J$ Model }

\author
{ 
Yasunari {\sc Tanuma}, Yukio {\sc Tanaka}, 
Masao {\sc Ogata},$^{1,2}$ and Satoshi {\sc Kashiwaya}$^{3}$
}

\inst
{
Graduate School of Science and Technology,
Niigata University, Ikarashi, Niigata 950-2102, Japan\\
$^{1}$ Department of Basic Science, Graduate School of Arts and Sciences,
University of Tokyo, 3-8-1 Komaba, Meguro-ku, Tokyo 153-0041, Japan\\
$^{2}$ Institute for Molecular Science Okazaki 444-0867, Japan \\
$^{3}$ Electrotechnical Laboratory, Tsukuba,
Ibaraki 305-0045, Japan
}

\recdate
{December 26, 1997}

\abst
{
Spatial dependencies of the pair potential
and the local density of states near the surfaces
of $d_{x^{2}-y^{2}}$-wave superconductors 
are studied theoretically.
The calculation is based on the $t$-$J$ model within a mean-field theory 
with Gutzwiller approximation.
Various types of surface geometries are considered.
Similar to our result in the extended Hubbard model,
it is found that the formation of zero-energy states  strongly
depends on the surface geometry. 
In addition to this feature, 
the zero-energy states give peak splitting for the (110) surfaces
when the super-exchange interaction $J$ is large.
This is due to the induced $s$-wave component near the surface.
The present result explains the microscopic origin of
the spontaneous time-reversal symmetry
breaking at the surfaces of high-$T_{c}$ superconductors. 
}

\kword
{
$t$-$J$ model, $d$-wave superconductor, Gutzwiller approximation,
scanning tunneling spectroscopy, zero-energy peaks,
spontaneous time-reversal symmetry breaking
}

\begin{document}
\sloppy
\maketitle

The $t$-$J$ model \cite{Zhang1} is believed to
explain the phase diagram of high-$T_{c}$
materials including the so-called pseudogap above $T_{c}$ and thus
is a realistic model for high-$T_{c}$ superconductors \cite{Fukuyama}.
Although there are many works regarding this model, 
the quasiparticle properties in nonuniform systems 
including  surfaces or interfaces 
are not clear at this stage except for those in the case with a vortex
\cite{Himeda}. 
Since $d_{x^{2} - y^{2}}$-wave pair potential 
is the most promising symmetry in the $t$-$J$ model 
for actual doping concentrations, 
we expect an interference effect of the quasiparticle 
due to the sign-change of the 
pair potential \cite{Harlin} at the surface.
\par
Recent theoretical works within quasiclassical approximation
revealed a novel interference effect peculiar to
the unconventional superconductivity
\cite{Kashiwaya1,Kashiwaya2,Hu,Tanaka1,Tanaka2,Matsumoto1,Barash}.
One of the interesting interference effects is the appearance
of zero-energy states (ZES) at the surface, which is due to
the sign-change of pair potential when the quasiparticles
are reflected at the surface \cite{Hu}.
This localized ZES manifests itself as a zero-bias
conductance peak (ZBCP) in tunneling experiments
\cite{Kashiwaya1,Covington}.
A recent experiment using a well-oriented (110) surface
detects a ZBCP \cite{Alff}.
However this ZBCP is often not detected experimentally,
which indicates that the presence of ZBCP strongly
depends on the surface quality. 
One of the main purposes of the present letter is to clarify
the origin of this dependence on the subtle difference of
the surface in the $t$-$J$ model.
\par
Another interesting feature is the splitting of ZBCP
which was observed in recent experiments
\cite{Covington,Kashiwaya3}.
This splitting is considered to be due to the coexistence
of a $s$-wave pair potential induced near the surface.
Till date, theories \cite{Fogel,Matsumoto2,Sigrist}
have assumed an additional attractive potential which
favors the $s$-wave component.
For example, Fogelstr\"{o}m {\em et al.} \cite{Fogel} suggested
electron-phonon interaction
as an origin of the $s$-wave component.
However, in this letter, we will show that the $t$-$J$ model
automatically induces the $s$-wave component near the surface
and thus gives a natural explanation of the splitting of ZBCP.
We discuss the doping and $J/t$ dependence of this splitting.
\par
Although the previous theories
\cite{Kashiwaya1,Kashiwaya2,Hu,Tanaka1,Tanaka2,Matsumoto1,Barash,
Fogel,Matsumoto2,Sigrist,Buchholtz,Nagato}
clarified some important properties of 
high-$T_{c}$ superconductors, 
they ignored several distinctive features characteristic to
high-$T_{c}$ materials: 
i) short coherence length, 
which invalidates the quasiclassical approximation 
and 
ii) strong correlation.
In order to deal with the effect of short coherence length, 
Tanuma {\em et al.} \cite{Tanuma1} 
developed local density of states (LDOS) theory
based on the extended 
Hubbard model beyond quasiclassical approximation. 
They clarified that LDOS of the 
$d_{x^{2}-y^{2}}$-wave superconductor is sensitive to the 
atomic structures near the surface. 
However, the strong correlation effect is not
sufficiently considered in their approach. \par
To consider this problem, 
we develop a theory of the $t$-$J$ model near the surface\cite{Ogata}. 
By using this model, we can naturally treat the $d$-wave superconductivity 
with short coherent length and strong correlation. 
The merit of studying the $t$-$J$ model is that we can systematically
investigate the doping or Fermi surface dependence which can be 
directly compared with the experimental results. 
In this letter, the LDOS of quasiparticles  
is obtained based on the self-consistently determined pair potential. 
Since it is difficult to treat the constraint in the $t$-$J$ model 
analytically, 
we apply the Gutzwiller approximation \cite{Zhang2}. 
The validity of this method was verified by comparing 
the variational energies of the bulk states  with 
those obtained in the variational Monte Carlo method \cite{Yokoyama}. 
After the Gutzwiller approximation, 
the spatial dependence of the pair potential is determined 
self-consistently within the mean-field approximation 
as in our previous work regarding vortex core \cite{Himeda}. 
\par
%
The $t$-$J$ model is given as \cite{Zhang1}
\begin{eqnarray}
\label{eq:01}
       {\cal H}= 
                 &-& t \sum_{\langle i,j \rangle,\sigma}
                 (\tilde{c}^{\dagger}_{i\sigma} \tilde{c}_{j\sigma}
                  + {\rm H.c.})
          + J\sum_{\langle i,j \rangle,\sigma} {\vct S}_{i} \cdot {\vct S}_{j} 
                  \nonumber \\ 
                 &-& \mu \sum_{i,\sigma} c_{i\sigma}^{\dagger} c_{i\sigma},
\end{eqnarray}
where $J$ and $\vct S_{i}$ are the super-exchange interaction and the spin-1/2
operator at the $i$-th site, respectively.
Here $\langle i,j \rangle$ stands for the summation over the
nearest-neighbor pairs.
The operator  $\tilde{c}_{i\sigma} = c_{i\sigma}P_{G}$
with the Gutzwiller projection operator
$P_{G}=\Pi_{i}(1 - n_{i\uparrow} n_{i\downarrow})$
excludes the double occupancy.
We use the Gutzwiller approximation  \cite{Zhang2}
in which the effect of the projection
is taken into account as statistical weights.
The expectation values are estimated as
\begin{eqnarray}
\label{eq:02}
\langle c^{\dagger}_{i\sigma} c_{j\sigma} \rangle
=g_{t}\langle c^{\dagger}_{i\sigma} c_{j\sigma} \rangle_{0},
\nonumber \\
\langle {\vct S}_{i} \cdot {\vct S}_{j} \rangle
=g_{s} \langle {\vct S}_{i} \cdot {\vct S}_{j} \rangle_{0},
\end{eqnarray}
where $\langle \cdots \rangle$ and $\langle \cdots \rangle_{0}$
represent the expectation values
in terms of a Gutzwiller-type variational wave function
$P_{G}|\Phi \rangle$ and a BCS wave function,
$|\Phi \rangle$, respectively.
The renormalized coefficients $g_{t}=2\delta/(1+\delta)$ and
$g_{s}=4/(1+\delta)^{2}$
with the hole concentration $\delta (=1-n)$
are determined from the probabilities of
involved configurations \cite{Zhang2,Yokoyama}.
Using this approximation, 
eq. (\ref{eq:01}) can be transformed into 
\begin{eqnarray}
\label{eq:03}
         {\cal H}_{\rm eff}=
                -t_{\rm eff}\sum_{\langle i,j \rangle,\sigma}
                (c^{\dagger}_{i\sigma} c_{j\sigma} + {\rm H.c.})
                +J_{\rm eff}\sum_{\langle i,j \rangle} {\vct S}_{i}
                \cdot {\vct S}_{j},
\end{eqnarray}
\begin{eqnarray}
\label{eq:04}
          t_{\rm eff} = g_{t}t, \quad 
         J_{\rm eff} = g_{s}J.
\end{eqnarray}
\par
In this letter, we consider various types of boundaries
as shown in Fig.~\ref{fig1}.
The index $m$ in Fig.~\ref{fig1}(d) denotes the period of zigzag
structures, and we will show that the LDOS
strongly depends on this period.
The case of $m=0$ [$m=1$] corresponds to a flat (100) [(110)]
surface shown in Fig.~\ref{fig1}(a) [(b)].
In the following, we discuss the cases with $m=0$, $1$ and $2$.
\par
For this model, we perform a mean-field approximation
with site-dependent pair potential $\Delta_{ij}$
and Hartree-Fock parameter $\xi_{ij\sigma}$,
\begin{eqnarray}
          \Delta_{ij} = \frac{3}{4}J_{\rm eff}
          \langle c_{i\uparrow} c_{j\downarrow} \rangle,
          \quad 
          \xi_{ij\sigma} = 
          \langle c_{i\sigma}^{\dagger} c_{j\sigma} \rangle.
\end{eqnarray}
Here we have assumed
$\xi_{ij\uparrow}=\xi_{ij\downarrow}=\xi_{ij}$.
For simplicity $\xi_{ij}$ and $\mu$ are fixed to the values $\xi_{0}$
and $\mu_{0}$ determined in the bulk without boundaries.
We use periodic boundary conditions in the $y$-direction and
open boundary conditions in the $x$-direction.
Furthermore, we assume that $\Delta_{ij}$ is translationally
invariant in the tangential direction along the surface.
Thus, the unit cell is $N_{L} \times 1$, with $N_{L}$ being
the number of sites in the $x$-direction.
After Fourier transformation,
the mean-field Hamiltonian becomes
\begin{eqnarray}
\label{eq:06}
      {\cal H}_{\rm eff} = \sum_{k_{y},i,j}
          \left(
                \begin{array}{cc}
                    C_{i\uparrow}^{\dagger}(k_{y}) & C_{i\downarrow}(-k_{y})
                 \end{array}
          \right)
          \left(
                 \begin{array}{cc}
                    \hat H_{ij} (k_{y}) & \hat \Delta_{ij} (k_{y}) \\
                    \hat \Delta^{\dagger}_{ji} (k_{y}) & -\hat H_{ji} (-k_{y})
                 \end{array}
          \right)
          \left(
                \begin{array}{c}
                    C_{j\uparrow}(k_{y}) \\
                    C_{j\downarrow}^{\dagger}(-k_{y})
                 \end{array}
          \right)
\end{eqnarray}
with
\begin{eqnarray}
\label{eq:07}
           \hat H_{ij}(k_{y}) &=&
                 - \sum_{\pm} [(t_{\rm eff} + \frac{3}{4}J_{\rm eff}\xi_{0})
                 (\delta_{i,j \pm 1} + e^{\mp ik_{y} a} \delta_{i,j \pm m})]
                 - \mu_{0} \delta_{i,j},    \\
\label{eq:08}
           \hat \Delta_{ij}(k_{y}) &=& \sum_{\pm}
                 [\Delta_{ij, x} \delta_{i,j \pm 1}
              + \Delta_{ij, y} e^{\mp ik_{y} a} \delta_{i,j \pm m}],
\end{eqnarray}
where 
$C_{j\sigma}(k_{y})$ is the Fourier transformed form 
of  $c_{j\sigma}$ with respect to the surface direction and
$j$ is now the site number in the $x$-direction ($j=1, \cdots, N_{L}$).
The wave vector $k_{y}$ changes from $-\pi/a$ to $\pi/a$. 
In the above,  $a$ is the lattice constant and we use $N_{L}=300$. 
The above Hamiltonian is diagonalized by Bogoliubov transformations 
\cite{Tachiki,Sato}
given as 
$C^{\dagger}_{j\uparrow}(k_{y})=\sum_{\nu}\gamma^{\dagger}_{\nu}(k_{y})
{\cal U}^{\ast}_{j,\nu}$, and 
$C_{i\downarrow}(-k_{y})=\sum_{\nu}\gamma^{\dagger}_{\nu}(k_{y})
{\cal U}^{\ast}_{N_{L}+i,\nu}$. 
The spatial dependence of the pair potential is determined
self-consistently as
\begin{eqnarray}
\label{eq:09}
  \Delta_{j \pm m,j, y} &=& \frac{3}{4} J_{\rm eff} \sum_{k_{y},\nu}
    {\cal U}_{j,\nu} {\cal U}^{\ast}_{N_{L}+j \pm m,\nu}
    \{1-f(E_{\nu}(k_{y}))\} e^{\pm ik_{y}a},     \\
\label{eq:10}
    \Delta_{j \pm 1,j, x} &=& \frac{3}{4} J_{\rm eff} \sum_{k_{y},\nu}
    {\cal U}_{j,\nu} {\cal U}^{\ast}_{N_{L}+j \pm 1,\nu}
    \{1-f(E_{\nu}(k_{y}))\}.
\end{eqnarray}
where $\Delta_{j \pm m,j,y}$
and $\Delta_{j \pm 1,j,x}$ are the pair potentials along the
$y$-axis and $x$-axis directions, respectively.
The above $f(E_{\nu}(k_{y}))$ denotes the Fermi distribution function.
\par
We solve the effective Hamiltonian in Eq. (\ref{eq:06}) by numerical
diagonalization and carry out an iteration
until the pair potentials  $\Delta_{i,j,x}$
and $\Delta_{i,j,y}$ are determined self-consistently. 
The obtained equations
are decomposed into real and imaginary parts
as
\begin{eqnarray}
         \Delta_{{\rm R},j,x(y)} &\equiv& {\rm Re}(\Delta_{ij,x(y)})/\Delta_{0},
         \nonumber \\
\Delta_{{\rm I},j,x(y)} &\equiv& {\rm Im}(\Delta_{ij,x(y)})/\Delta_{0}.
\end{eqnarray}
\par
Figure~\ref{fig2}(b) shows the calculated results of the spatial dependence of
the pair potential for the flat (110) surface for $J/t=0.4$ and $\delta=0.15$. 
For this geometry, 
$\Delta_{{\rm R},j,x(y)}$ 
and  $\Delta_{{\rm I},j,x(y)}$ can be regarded as 
$d$-wave and extended $s$-wave components of the pair potential,
respectively. 
The quantity $\Delta_{{\rm R},j,x(y)}$
is suppressed at the surface and increases
monotonically as we approach the middle of the lattice.
This behavior is consistent with the quasiclassical theory \cite{Nagato}.
The extended $s$-wave component is induced near the surface,
whose magnitude is about 30\%
relative to the bulk $d$-wave component.
We find an atomic-scale spatial oscillation of the $s$-wave component,
which is completely neglected in the quasiclassical approximation.
\par
Using the self-consistently determined  pair potential,
we calculate the LDOS at every site.
In order to compare our theory with
scanning tunneling microscopy (STM) experiments,
we assume that the STM tip is metallic with a flat density of states,
and that the tunneling probability is finite
only for the nearest site from the tip. 
The LDOS at $i$-th site is given as,
\begin{eqnarray}
\label{eq:11}
   \rho_{i} & \sim &
   \int ^{\infty}_{-\infty}d\omega \rho_{i}(\omega)
   {\rm sech}^{2}(\frac{\omega + E}{2k_{B}T}), \\
   \rho_{i}(\omega)
      & = & -\frac{2}{\pi} {\rm Im} \sum_{k}G^{R}_{i}(k,\omega) 
      \nonumber \\
      & = &2 \sum_{k} \sum_{\nu} |{\cal U}_{i,\nu}|^{2}
      \delta(\omega-E_{\nu}(\vct k))
\end{eqnarray} 
where $G^{R}_{i}(k,\omega)$ is the Fourier component of the retarded
Green's function with energy $\omega$. 
In the actual STM experiments, since the magnitude of the 
transparency between the tip and surface is small, 
the tunneling conductance converges to 
the normalized LDOS
\begin{eqnarray}
\label{eq:13}
   \bar{\rho}(E) =
   \frac{
   \displaystyle{
   \int ^{\infty}_{-\infty}d\omega \rho_{i,S}(\omega)
   {\rm sech}^{2}(\frac{\omega + E}{2k_{B}T})}}
   {
   \displaystyle{
   \int ^{\infty}_{-\infty}d\omega \rho_{N}(\omega)
   {\rm sech}^{2}(\frac{\omega + 2\Delta_{0}}{2k_{B}T})}},
\end{eqnarray}
at low temperatures \cite{Kashiwaya2}, 
where $\rho_{i,S}(\omega)$ denotes the LDOS 
in the superconducting state 
and $\rho_{N}(\omega)$ 
denotes the LDOS in the normal state.
In this letter,  $\rho_{N}(\omega)$ is obtained from
the LDOS at the $N_{L}/2$-th site away from the boundary. 
\par
%
%
Figure~\ref{fig2}(c) shows the calculated LDOS
for various sites near the flat (110) surface.
This zero-energy peak (ZEP) is  the manifestation of ZES,
which are formed due to the sign-change
of the $d_{x^{2}-y^{2}}$-wave pair potential.
A remarkable difference between the present results and those
based on the quasiclassical theory 
is seen in the oscillatory behaviors of the LDOS
\cite{Matsumoto1}.
This oscillation can be regarded as the 
Friedel oscillation, the period of which is 
the inverse of the Fermi momentum. 
Furthermore, we find that the ZEP of the LDOS
is split into two at all sites
near the surface (see Fig.~\ref{fig2}(c)). 
The splitting of the ZBCP in tunneling spectra 
is also obtained in the quasiclassical approximation
\cite{Matsumoto2,Fogel}. 
Its origin is the $s$-wave component induced near the surface 
which blocks the motion of quasiparticles 
near the (110) surface \cite{Himeda,Matsumoto2}.
In effect, the splitting is not visible when $J/t$ is decreased
and the amplitude of the induced imaginary component is small.
On the other hand, 
when $\delta$ is increased, 
the splitting becomes larger. 
This is because
$\Delta_{{\rm R},j,x(y)}$ is reduced in magnitude with the increase of $\delta$,
while $\Delta_{{\rm I},j,x(y)}$ is insensitive to 
the change of $\delta$. 
Recently, experimental observations of 
the peak splitting are reported 
in the tunneling spectroscopy of 
high-$T_{c}$ superconductors \cite{Covington,Kashiwaya3}. 
This indicates that the $s$-wave component is induced
near the sample surface and the spontaneous time-reversal
symmetry breaking actually takes place in high-$T_{c}$
superconductors. \par
We also study the flat (100) surface shown
in Fig.~\ref{fig1}(a).
In this case, $\Delta_{{\rm I},j,x(y)}$ 
is not induced near the surface.
Since the quasiparticles do not feel the sign-change of the
pair potential at the (100) surface,
the ZEP do not appear.
\par
Next, we discuss the case of a $1 \times 2$ zigzag surface 
(see Fig.~\ref{fig3}(a)). 
The obtained pair potential has complex
spatial dependence 
as compared to the cases with (100) and (110) surfaces. 
The quantity $\Delta_{{\rm I},j,x(y)}$ 
is not induced as shown in Fig.~\ref{fig3}(b). 
On the other hand, $\Delta_{{\rm R},j,x(y)}$ has 
an oscillation and behaves like 
a $d_{x^{2}-y^{2}}$ pair potential far away from the boundary.
The complex spatial dependencies of the pair potential
reflect on the LDOS as an anomalous structure 
with many dips and peaks (see Fig.~\ref{fig3}(c)). 
It should be noted that the 
ZES are not formed near the surface. 
A similar property was also obtained on the $1 \times 2$ zigzag 
surface in the extended Hubbard model \cite{Tanuma2}. 
The reason for the absence of ZEP is as follows.
The wave function of ZES spatially 
oscillates with the period of the 
inverse of the Fermi momentum as discussed 
in the (110) surface \cite{Fogel,Matsumoto2,Sigrist}. 
In the underdoped region,
since the Fermi surface is nearly square, 
the period of the oscillation of the  wave function is 
roughly coincident with $2a$. 
Consequently, the node and the antinode   appear 
alternatively.
However, for the $1 \times 2$ zigzag structure the phase of
node and antinode does not coincide.
This disappearance of the ZES can be regarded 
as an interference effect 
of the standing-wave, which cannot be explained by means of
the quasiclassical theory. 
\par
%
%
In this letter, we have investigated the LDOS
near the surfaces of the $d_{x^{2}-y^{2}}$-wave superconductor
based on the $t$-$J$ model within the Gutzwiller approximation.
In the present calculation, the non-local feature of the pair potential
and the atomic-scale geometry of the surface are explicitly taken into
account. 
The present result gives a microscopic basis 
for the spontaneous time-reversal
symmetry breaking in $d_{x^{2}-y^{2}}$-wave superconductors where 
the $s$-wave component is induced as the $d+is$ state near the surface. 
It is clarified that when  the amplitude of the induced $s$-wave component is enhanced
with the increase of the magnitude  of $J$, the 
ZEP of LDOS is split into two.  
\par
This work is supported by a Grant-in-Aid for Scientific
Research in Priority Areas
``Anomalous metallic state near the Mott transition''
and ``Nissan Science Foundation''.
The computational aspect of this work has been performed at the
facilities of the Supercomputer Center, Institute for Solid State Physics,
University of Tokyo and the Computer Center, Institute for Molecular
Science, Okazaki National Research Institute.
%

\newpage
\begin{figure}
%
\caption{Schematic illustration corresponding to $1 \times m $
zigzag surface : (a) a flat (100) surface ($m=0$),
(b) a flat (110) surface ($m=1$),
(c) a $1 \times 2 $ zigzag surface ($m=2$)
and
(d) a $1 \times m $ zigzag surface.}
\label{fig1}
\end{figure}
\begin{figure}
%
\caption{(a) Schematic illustration of a flat (110) surface,
(b) the spatial dependence of the normalized pair potential
and (c) the local density of states for $J/t=0.4$ and $\delta=0.15$
($\Delta_{0}/t=0.105$).}
\label{fig2}
\end{figure}
\begin{figure}
%
\caption{(a) Schematic illustration of a $1 \times 2$ zigzag surface,
(b) the spatial dependence of the normalized pair potential
and (c) the local density of states for $J/t=0.2$ and $\delta=0.1$
($\Delta_{0}/t=0.054$).}
\label{fig3}
\end{figure}
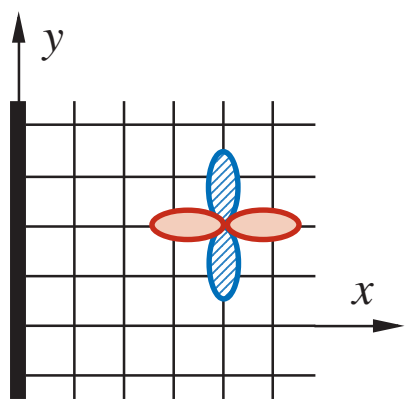
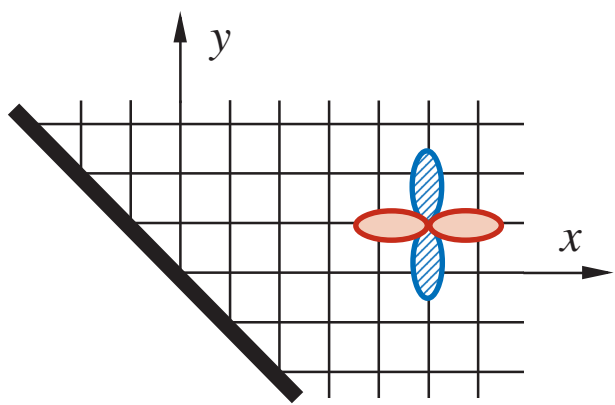
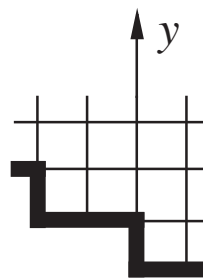
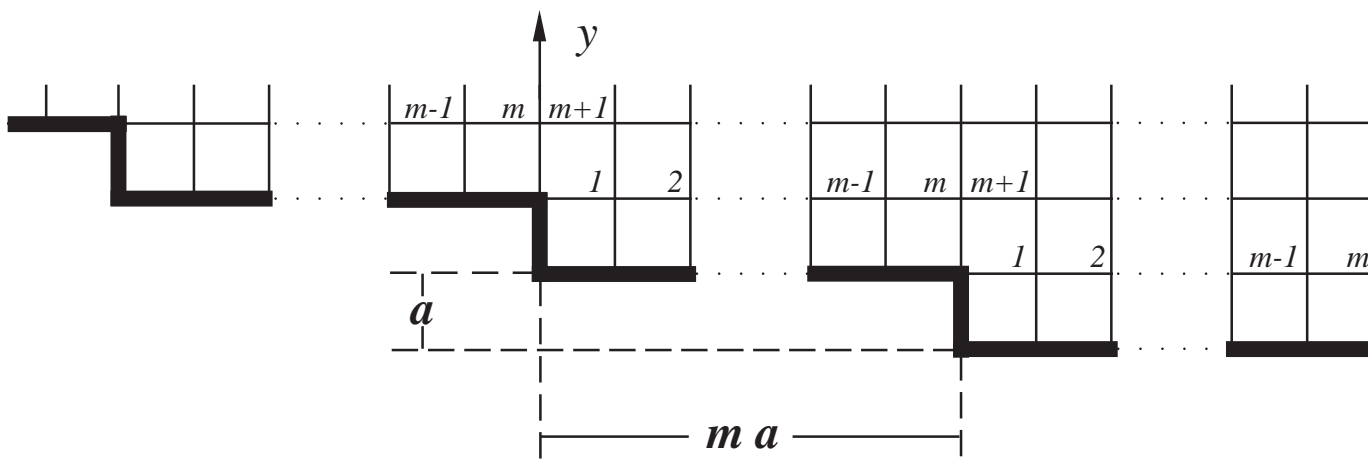
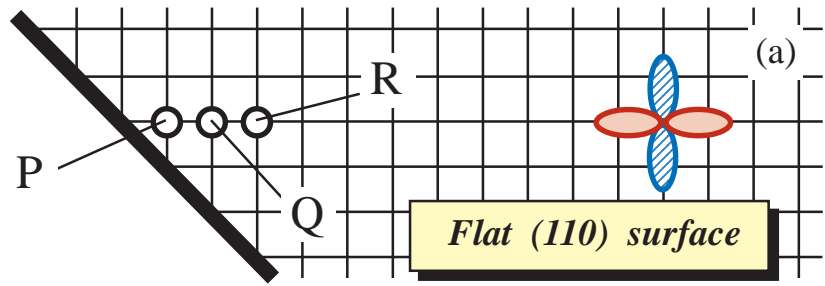
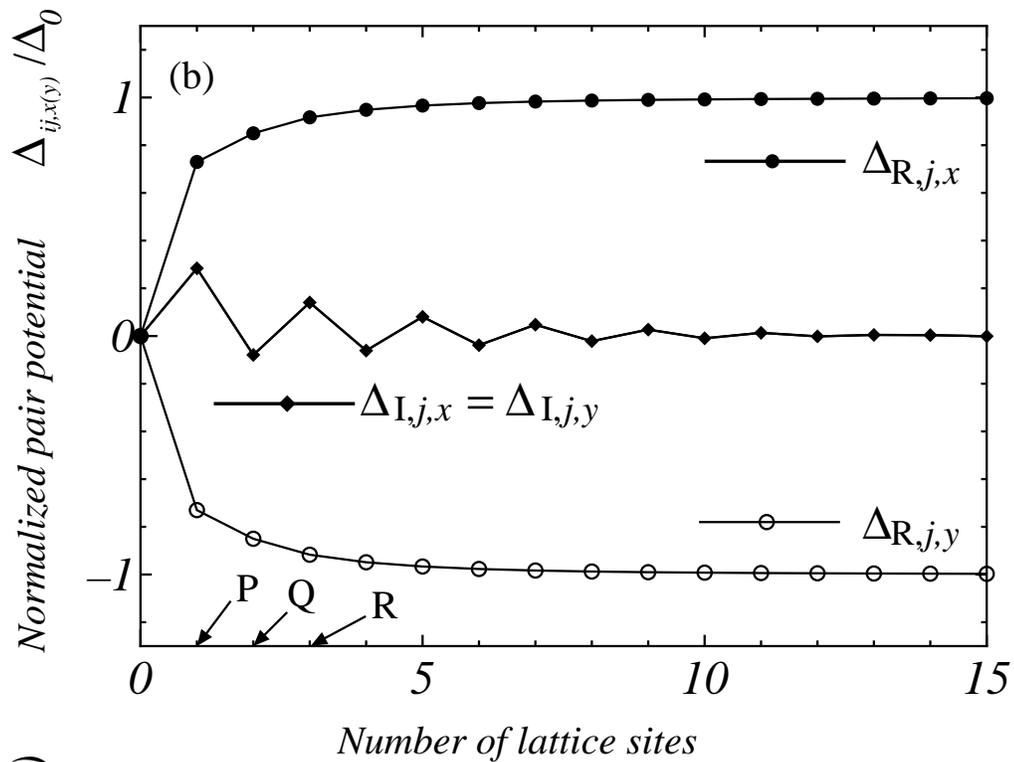
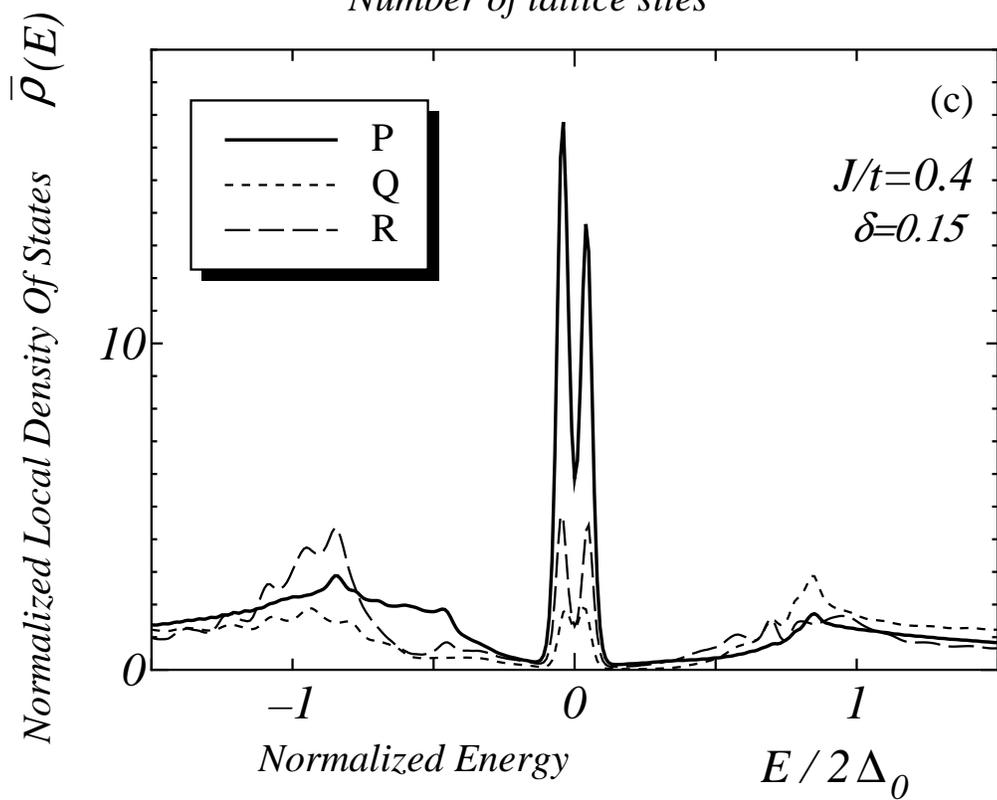
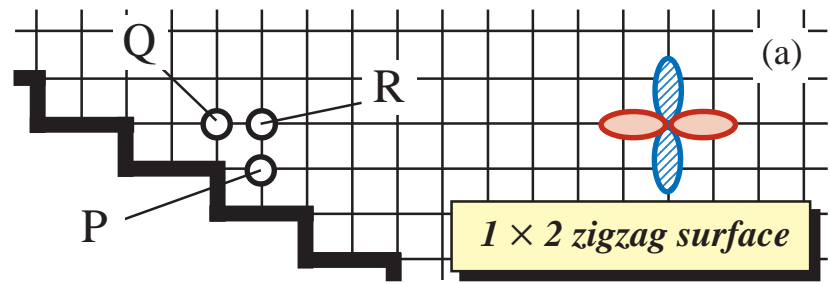
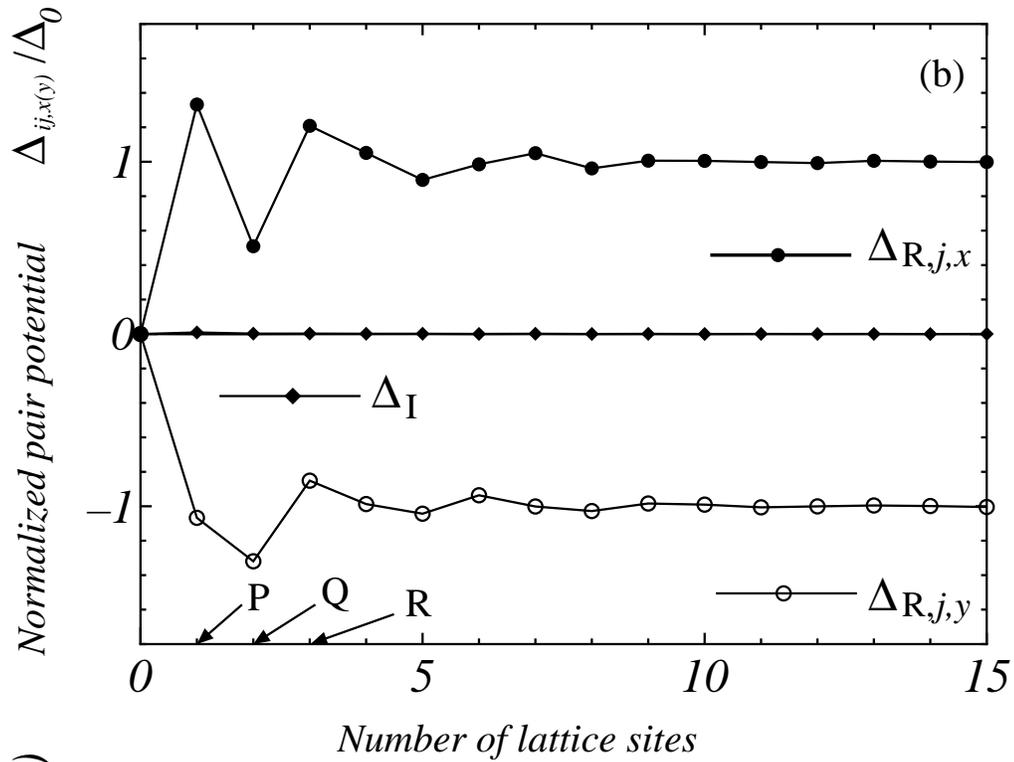
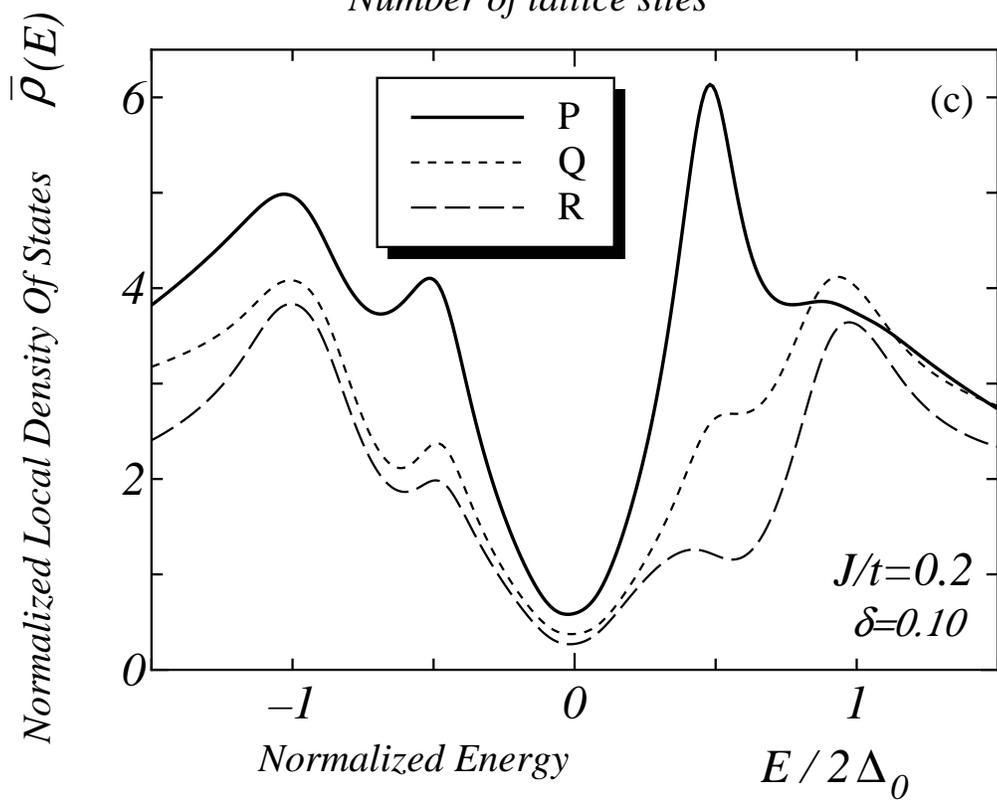
\end{document}